\documentclass[%
reprint,
amsmath,amssymb,
prapplied,superscriptaddress
]{revtex4-2}
\usepackage[utf8]{inputenc}
\usepackage{graphicx}% Include figure files
\usepackage{dcolumn}% Align table columns on decimal point
\usepackage{bm}% bold math
\usepackage{mathrsfs}
\usepackage{amsfonts}
\usepackage{physics}
\usepackage{xcolor}

\DeclareMathOperator{\diag}{diag}

\begin{document}

%\title{Quantum-Optimal Sensing of a Parameter Locked in Distributed Phase Functions}
%\title{Quantum-enhanced sensing of a phase distributed to multiple functions}
\title{Entanglement-Enhanced Estimation of a Parameter Embedded in Multiple Phases}

\author{Michael R. Grace}
\email{michaelgrace@email.arizona.edu}
\affiliation{James C. Wyant College of Optical Sciences, University of Arizona, Tucson, AZ 85721, USA}
\author{Christos N. Gagatsos}
\affiliation{James C. Wyant College of Optical Sciences, University of Arizona, Tucson, AZ 85721, USA}
\author{Saikat Guha}
\affiliation{James C. Wyant College of Optical Sciences, University of Arizona, Tucson, AZ 85721, USA}

\begin{abstract}
	Quantum-enhanced sensing promises to improve the performance of sensing tasks using non-classical probes and measurements that require far fewer scene-modulated photons than the best classical schemes, thereby granting previously-inaccessible information about a wide range of physical systems. We propose a generalized distributed sensing framework that uses an entangled quantum probe to estimate a scene-parameter encoded within an array of phases, with a functional dependence on that parameter determined by the physics of the actual system. The receiver uses a laser light source enhanced by quantum-entangled multi-partite squeezed-vacuum light to probe the phases and thereby estimate the desired scene-parameter. The entanglement suppresses the collective quantum vacuum noise across the phase array. We report simple analytical expressions for the Cram\'er Rao bound that depend only on the optical probes and the physical model of the measured system, and we show that our structured receiver asymptotically saturates the quantum Cram\'er-Rao bound in the lossless case. Our approach enables Heisenberg limited precision in estimating a scene-parameter with respect to total probe energy, as well as with respect to the number of modulated phases. Furthermore, we study the impact of uniform loss in our system and examine the behavior of both the quantum and the classical Cram\'er-Rao bounds. We apply our framework to examples as diverse as radio-frequency phased-array directional radar, beam-displacement tracking for atomic-force microscopy, and fiber-based temperature gradiometry.
\end{abstract}

\maketitle

\section{Introduction}
Quantum phenomena are now known to be powerful and viable tools to enhance estimation precision in diverse fields, e.g., astronomy~\cite{Khab2019}, general relativity~\cite{Bruschi2014,Branford2018,LIGO1,LIGO2}, models for quantum-to-classical transition~\cite{Branford2019}, microscopy~\cite{Bisketzi2019}, and optical imaging~\cite{Humphreys2013,Gagatsos2016,Knott2016}. Quantum-enhanced estimation in sensing applications, which arguably comprises the nearest-term realizable quantum technologies of practical importance, promises an improvement the sensitivity of estimating an unknown parameter of the physical system being probed. In an idealized sensing context, this improvement takes the form of an improved scaling of estimation variance with probe power, known as Heisenberg scaling. Moreover, this Heisenberg scaling for sensitivity can be obtained using {\em Gaussian} quantum states of light (that can be generated using lasers, linear optics, and squeezed light, e.g., produced using parametric amplifiers) and {\em Gaussian} measurements (i.e., homodyne and heterodyne detection).  This is an especially interesting point, since Gaussian resources alone do {\em not} suffice to perform a variety of other quantum information tasks~\cite{Bartlett2002,Eisert2002-ys,Fiurasek2002-ip,Giedke2002-kz,Cerf2005-ck,Takeoka2008-jh,Wittmann2010-lm,Niset2009-sp,Takeoka2014-nb}. 

In the context of distributed, or networked, quantum sensing, entangled Gaussian states have been shown to yield additional, significant advantages over separable quantum-enabled probes, for which the states of individual optical modes can be independently characterized ~\cite{Ge2018,Proctor2018,Zhuang2018-zu,Gatto2019,Guo2019,Grace2020,Oh2020}. These entangled probes have the advantage of being generated from separable Gaussian states and Gaussian unitary operations \cite{Ge2018}, both of which can be readily analyzed~\cite{Weedbrook2012,Ferraro2005} and realized experimentally~\cite{Chen2014-kf,Takanashi2019-ab}.

One widely applicable scenario for distributed quantum sensing is as follows: a quantitative parameter of interest $x$ modulates a series of $M$ optical phase delays in $M$ optical modes with non-symmetric but known functional relationships to the parameter, and the parameter must be estimated using user-controlled probes and measurements. This problem statement can be used to model various practical photonic sensing tasks, some specific examples of which have been recently studied. These include the estimation of the angle of incidence, or other attributes of a radio-frequency (RF) wave upon an array of sensor pixels where each pixel is a phase modulator that is read out optically~\cite{Xia2020}; estimating a small transverse displacement of an optical beam in an atomic-force microscope (AFM)~\cite{Qi2018}; estimation of a small angular velocity of rotation of a Sagnac-based fiber-optical gyroscope (FOG)~\cite{Grace2020}; and estimating material defects with a fiber-based temperature gradiometer. 
%In the above examples, $M$ respectively would refer to: the number of sensor pixels of the RF photonic antenna; the number of stacked fiber loops in a FOG; and the number of orthogonal spatial modes in the probe's free-space propagation geometry. Likewise, the scene-parameter of interest $x$ could refer to: the angle of incidence of the impinging RF field; the radians-per-second rotation experienced by the FOG; and the longitudinal displacement of a nano-cantilever of an AFM. This setting also models a , a doppler vibrometer, and more. 
General results for variants of this scenario have been found in recent work focusing on measurements that perfectly achieve the ultimate bounds on estimation precision ~\cite{Matsubara2019} and handle imperfections in prior knowledge of the parameter~\cite{Gramegna2021,Gramegna2021a}. 
%~\footnote{While writing this paper, we came across concurrent research that analyzed a closely-related model~\cite{Gramegna2020a,Gramegna2020b}. Our paper differs from and complements these papers in the following ways. We calculate the QFI to evaluate the fundamental limit on precision attainable with {\em any} receiver design. Further, our optical probe is different in that it includes a strong coherent-state (laser-light) source, so that the entangled squeezed light only contributes to a small portion of the total mean photon number budget. Designs where squeezed light contributes {\em all} the probe power require much more squeezing to achieve a given performance, which would be harder to generate. Finally, we do not use an adaptive measurement since we assume our scene-parameter $x$ is very close to a pre-estimate or {\em a priori} known value $x_0$, i.e., $|x-x_0|\ll 1$. Our choices were based on the practical photonic sensors described in the Introduction, and with experimental feasibility of implementation in mind.}

Here, we propose a framework for quantum sensing that combines the fundamental Heisenberg scaling advantage offered by entangled Gaussian probe states with practicality in analysis and implementation for real-world applications. The two estimation theory tools we use are the quantum Fisher information (QFI) $H_x$ and the classical Fisher information (CFI) $I_x$. Both give lower bounds to the mean squared error (MSE) $\langle (\hat{x}-x)^2\rangle$ for unbiased estimators $\hat{x}$ (i.e., $\langle \hat{x} \rangle = x$), via the classical and quantum Cram\'er-Rao bounds: $\langle (\hat{x}-x)^2\rangle \geq I_x^{-1} \geq H_x^{-1}$. We consider the lossless case to derive the fundamental optimal quantum performance and we also study the effects of uniform pure loss.  Applications of our model reach beyond photonic sensors, and could serve as the foundation for more general distributed quantum sensing tasks with applications to quantum process tomography of quantum computers and quantum network tomography.

\section{Physical Model and Estimation Task}
\begin{figure}
	\centering
	\includegraphics[width=\columnwidth]{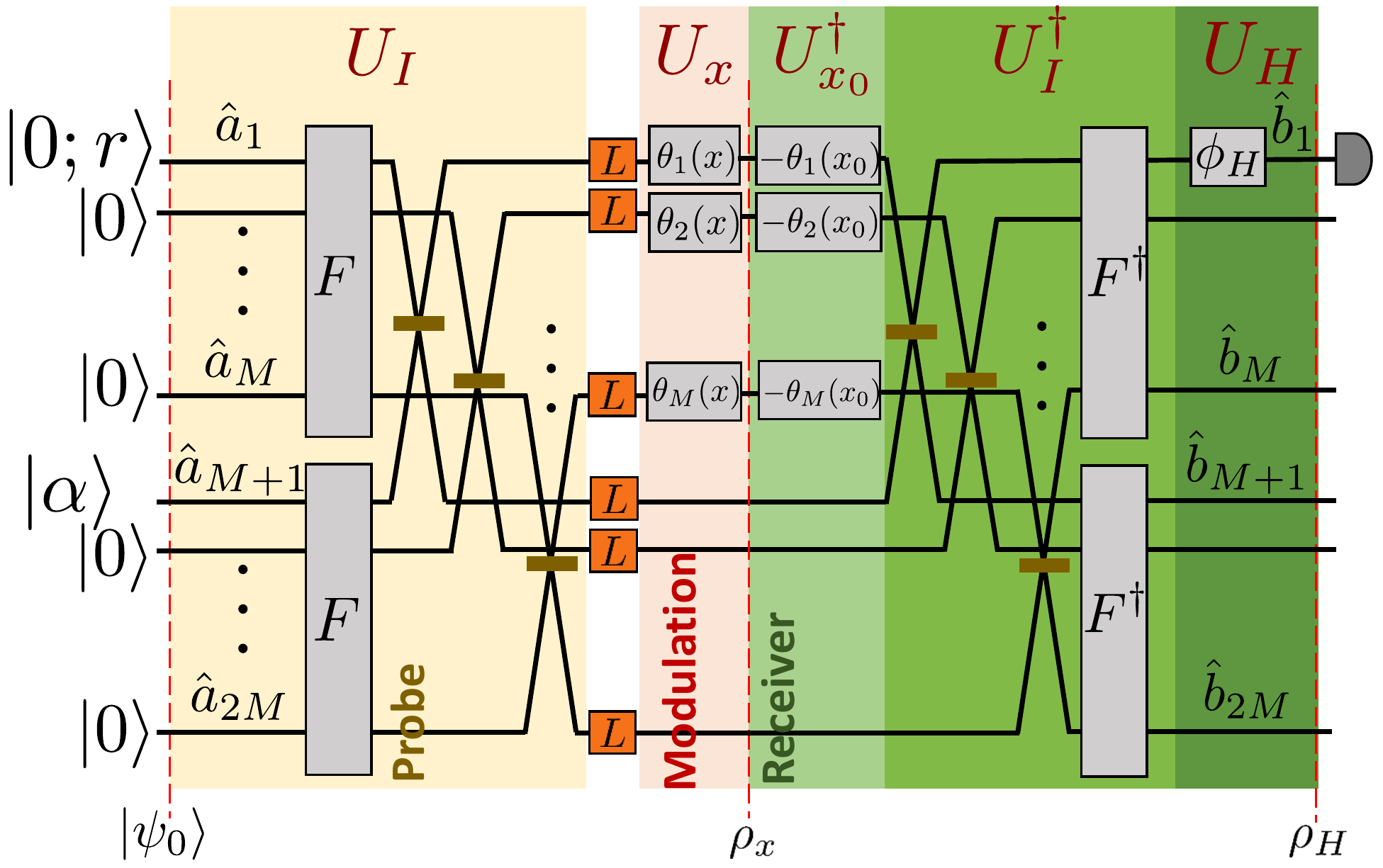}
	\caption{Sensing of a single parameter of interest $x$ embedded in phase functions $\theta_m(x)$ modulating $M$ MZIs. Each two-mode MZI is probed with a coherent state and one mode of an $M$-mode-entangled squeezed vacuum state. $F$, the Fourier gate, is an $M$-mode linear-optical interferometer, and the detector at the output of the circuit is a homodyne detector. L denotes the pure loss channel which is modeled as a beam splitter with transmissivity $\tau$ whose lower input mode is set to vacuum.  A local oscillator mode for homodyne detection is implied but not shown for simplicity.}
	\label{fig:SensorDiagram}
\end{figure}

In this work, we consider a scenario shown in Fig.~\ref{fig:SensorDiagram}: a single scalar scene-parameter $x$ modulates phases $\theta_m(x)$, $1 \le m \le M$, in an array of $M$ Mach-Zehnder Interferometers (MZIs), each with a pair of input and output modes. The state preparation circuit and the phase modulation can be described by the passive unitary transformations
\begin{eqnarray}
	\label{eq:UI} U_{\rm I} &=& F_{M\cross M}\otimes F_{2\cross2} \\
	\label{eq:Ux} U_x &=& D\big(\vec{\theta}(x)\big) \oplus\mathcal{I}_{M\cross M},
\end{eqnarray}  
where $F$ signifies a linear-optic Fourier gate, $D\big(\vec{\theta}(x)\big)$ is an $M \cross M$ diagonal matrix with entries $\theta_m(x)$, and $\mathcal{I}$ is an identity matrix. In addition, the $2M$ optical modes of the circuit each undergo identical pure loss channels with optical transmissivity $0\leq\tau\leq1$, imposing mode-symmetric loss on the probe state before modulation by the parameter-dependent phases (e.g., induced during propagation to spatially distributed sensors). Given a total mean photon number budget $N$, one is tasked to design a $2M$-mode (quantum) optical probe and an associated receiver to minimize the variance of the estimate of $x$. The probe we consider consists of a single-mode squeezed vacuum (SV) $\ket{0;r}$ with squeezing parameter $r>0$ and a (laser light) coherent state $\ket{\alpha}$ with complex amplitude $\alpha \in \mathbb{C}$. Each of these two inputs is equally mixed with $M-1$ vacuum modes on balanced Fourier gates. The output of the first Fourier gate is an $M$-mode entangled continuous-variable (CV) state~\cite{Zhuang2018-zu}, and the output of the second is $M$ identical (product) coherent states, $\ket*{\alpha/\sqrt{M}}$. Our scheme applies a linear-optical receiver circuit to the $M$ modulated modes followed by single-mode homodyne detection. We will show that when the parameter $x$ is known to be close to some value $x_0$, an appropriate recombining operation involves a set of phase shifts $U_{x_0}$, the inverse unitary $U_{\rm I}^{\dagger}$, and a single-mode phase shift $U_{\rm H}$.

\section{Quantum Fisher Information}
The QFI $H_x$ quantifies optimal precision in estimating $x$ by {\em any} receiver. As $H_x$ is directly a function of $\rho_x$, it is dependent on the choices made for the input state $\ket{\psi_0}$ and the preparation circuit $U_{\rm I}$. In the quantum metrology literature~\cite{Pirandola2018-gw}, it has been shown for various settings that in the absence of loss and noise, quantum optical probes can attain the so-called Heisenberg scaling $H_x = O(N^2)$, where $N$ is the total photon-unit energy in the probe field. In contrast, sensing schemes using classical probes can only achieve $H_x = O(N)$ at best. If the CFI $I_x$ for a specific quantum probe and a receiver scales as $O(N^2)$, that system design achieves Heisenberg scaling with probe energy, while a CFI $I_x=H_x$ indicates a quantum optimal receiver for a given probe.

The QFI leading to a quantum Cr\'amer-Rao bound on the estimation of the parameter $x$ can be expressed as follows ($\partial_x \equiv \partial/\partial x$):
\begin{equation}
	\label{eq:QFI_Jacobian}	H_x =\sum_{k,l=1}^{M} \partial_x \theta_k (x) \partial_x \theta_l (x) H_{kl}.
\end{equation}
where $H_{kl}$ are the indexed matrix elements of the quantum Fisher information matrix for multiple parameter estimation of the $M$ phases $\theta_m(x)$. Since QFI is agnostic of the chosen receiver, we compute the $H_{kl}$ from the intermediate, generally mixed, state $\rho_x$ that is found immediately after modulation by the parameter-dependent phases. 

Since the Gaussian input state $\ket{\psi_0}$ is transformed to $\rho_x$ solely by Gaussian transformations, the state $\rho_x$ will also be Gaussian. The QFI matrix elements $H_{kl}$ can then be computed from the displacement vector $\vec{d}_x$ and covariance matrix (CM) $V_x$ of the state $\rho_x$. The displacement vector and CM of the probe state shown in Fig.~\ref{fig:SensorDiagram} are \footnote{Our conventions are that the elements of first moments and CMs are indexed according to $\vec{r}=\{\vec{q},\vec{p}\}$, where $\vec{q}$ and $\vec{p}$ are the $2M$ real and imaginary mode quadratures, and $\hbar\equiv 1$}
\begin{equation}
	\vec{d}_{0,m}=
	\begin{cases}
		q_0,&\, m=M+1\\
		p_0,&\, m=3M+1\\
		0,&\, \textrm{otherwise}
	\end{cases}
	\label{eq:d0}
\end{equation}
where $\alpha=(q_0+ip_0)/\sqrt{2}$ and $V_0=\frac{1}{2}\diag(\vec{v}_0)$, where
\begin{equation}
	\vec{v}_{0,m}=
	\begin{cases}
		e^{2r}&\, m=1\\
		e^{-2r}&\, m=2M+1\\
		1,&\, \textrm{otherwise}.
	\end{cases}
	\label{eq:V0}
\end{equation}
In general, the displacement vector and CM of a Gaussian state transform evolve through a Gaussian unitary $U$ as $\vec{d}_{\rm out}=S\vec{d}_{\rm in}$ and $V_{\rm out}=SV_{\rm in}S^T$, where $S$ is the symplectic matrix satisfying
\begin{equation}
	\label{eq:symplectic}	S = \mathcal{I}_{2\times 2} \otimes \Re\{U\} -\Omega \otimes \Im\{U\},
\end{equation}
with $\Omega=\text{antidiag}(1,-1)$. Furthermore, a generic displacement vector and CM evolve through a symmetric pure loss channel via
\begin{eqnarray}
	\label{eq:LossDisplacement} \vec{d}_{\rm out} &=& X_{\tau}\vec{d}_{\rm in}\\
	\label{eq:LossCovariance} V_{\rm out} &=& X_{\tau}V_{\rm in}X^{T}_{\tau} + Y_{\tau},
\end{eqnarray}
where $X_{\tau}=\sqrt{\tau} \mathcal{I}$, $Y_{\tau}=[(1-\tau)/2] \mathcal{I}$, $\mathcal{I}$ is the $4M\times4M$ identity matrix, and $0\leq\tau\leq 1$ is the transmittance the channel acting identically on each mode.

After some calculations (Appendix~\ref{apx:QFI-Loss}), we find that
\begin{equation}
	H_{kl} = \frac{2}{M}\left(\sinh^2 s+\tau |\alpha|^2+\bar{N}_1\cosh 2s\right) \delta_{kl} +\frac{h}{M^2},
	\label{eq:QFI_phases}
\end{equation}
where the mean thermal photon number
\begin{equation}
	\label{eq:ThermalPhotons}	\bar{N}_1 =  \sqrt{\tau (1-\tau)\sinh^2 r+\frac{1}{4}}-\frac{1}{2}
\end{equation}
and the reduced squeezing parameter
\begin{equation}
	\label{eq:squeezing} s=\frac{1}{4} \ln \left[\frac{1+(e^{2r}-1)\tau}{1+(e^{-2r}-1)\tau}\right],
\end{equation}
are given as functions of the input squeezing $r$ and the symmetric transmittance $\tau$, and where
\begin{equation}
	\begin{split}
		h=& \frac{1}{4 (4 \bar{N}_1^3+6 \bar{N}_1^2+4 \bar{N}_1+1)} \times\\
		& \Big\{ 8 \tau |\alpha|^2 \big[2 \bar{N}_1 (\bar{N}_1+1)+1\big] (\sinh^2 s-\bar{N}_1 ) +\\
		& + 2 (\alpha^2+\alpha^{*2}) \big[2 \bar{N}_1 (\bar{N}_1+1)+1\big] \tau  \sinh 2s+\\
		& + (2 \bar{N}_1+1)^3 \cosh 4s-\\
		&2 \big[2 \bar{N}_1 (\bar{N}_1+1)+1\big] (2 \bar{N}_1+1)^2 \cosh 2s+2 \bar{N}_1+1\Big\}.
	\end{split}
	\label{eq:h}
\end{equation}

Therefore, the QFI for the parameter $x$ under uniform pure loss is found by utilizing Eq. \eqref{eq:h} and inserting Eq. \eqref{eq:QFI_phases} into Eq. \eqref{eq:QFI_Jacobian} and takes the form,
\begin{eqnarray}
	\nonumber	H_x &=&2 \left(\sinh^2 s+\tau |\alpha|^2+\bar{N}_1\cosh 2s\right) \langle\partial_x \theta^2(x)\rangle\\
	\label{eq:QFI_Loss}	&&+h \langle \partial \theta(x)\rangle^2,
\end{eqnarray}
where $\expval*{\partial\theta(x)}=(1/M)\sum_{m=1}^M\partial_x\theta_m(x)$, and $\expval*{\partial\theta^2(x)}=(1/M)\sum_{m=1}^M\big(\partial_x\theta_m(x)\big)^2$.

By choosing $r>0\Rightarrow s>0$, the choice of $\alpha$ that maximizes the QFI of Eq. \eqref{eq:QFI_Loss} is $\alpha=\alpha^*$. This can be seen by noting that the only dependence of Eq. \eqref{eq:QFI_Loss} on an $\alpha$ which is not in a modulo, is in Eq. \eqref{eq:h}. There for $s>0$, the prefactor of $\alpha^2+\alpha^*$ is positive and therefore the choice $\alpha=\alpha^*$ is optimal.

In the absence of loss (i.e., $\tau=1$), we find that as a function of the mean photon numbers $N_s=\sinh^2r$ and $N_v=\abs{\alpha}^2$ of the input SV state and coherent state, the QFI of Eq.~\eqref{eq:QFI_Loss} reduces to
\begin{eqnarray}					
	\nonumber		H_{x}&=&\expval*{\partial\theta(x)}^2\Big[N_v\big(\sqrt{N_s}+\sqrt{N_s+1}\big)^2\\
	\nonumber &&+2N_s(N_s+1)-N-4p_0^2\sqrt{N_s(N_s+1)}\Big]\\
	&&+2\expval*{\partial\theta^2(x)}N,
	\label{eq:QFI_result}
\end{eqnarray}
where $N=N_v+N_s$. The lossless QFI is clearly optimized by choosing $\alpha\in\mathbb{R}$, a property inherited from Eq. \eqref{eq:QFI_Loss} which is valid for all $0\leq\tau\leq 1$ .

By inspection, the first two terms of Eq.~\eqref{eq:QFI_result} scale quadratically with input energy as $N_v\gg1$ and $N_s\gg 1$, providing the Heisenberg scaling advantage available with quantum probes; the first term depends on the energy contribution from both input sources, while the second depends on the energy from the SV source.
%The commutation relations between normal-ordered and symmetrically-ordered moments becomes significant when working with the Wigner function, as explained in Appendix A. 

\section{Performance Evaluation of Proposed Receiver}
Let us assume that $x$ is known to be close to a value $x_0$, i.e., $|x-x_0|\ll 1$, due to either \emph{a priori} information or a preliminary estimate of $x$. Under this condition, applying the conjugate phases $-\theta_m(x_0)$ to the state $\rho_x$ followed by a sequence of 50-50 beamsplitters (i.e., the second beamsplitters of the MZIs) and inverse Fourier gates $F^{\dagger}$ efficiently recombines the information-bearing light to one desired output mode $\hat{b}_1$  (Fig.~\ref{fig:SensorDiagram}). The phase $\phi_H$ controls which field quadrature is measured via homodyne detection. The phases $-\theta_m(x_0)$ could be unitarily evolved to a different set of phases $-\tilde{\theta}_m(x_0)$ that are applied {\em after} the second set of beamsplitters; this configuration maintains the physical sensor's natural mathematical description  as an array of $M$ MZIs \cite{Xia2020,Grace2020,Qi2018} while not changing the CFI. If prior knowledge is not available of the functional form of the phases $\theta_m(x)$ \cite{Zhuang2019} or of the parameter $x$ itself \cite{Gramegna2021,Gramegna2021a}, the linear recombination of light must be determined via an adaptive strategy. 

We begin by assuming lossless optical sensing (i.e., $\tau=1$). The $2M$-mode output state $\rho_{\rm H}$ is a pure state and is determined by the input state $\ket{\psi_0}$ and the full system unitary $U(x)=U_{\rm H} U_{\rm I}^{\dagger} U_{x_0}^{\dagger} U_x U_{\rm I}$, where $U_{\rm H} = {\text{diag}}(e^{i\phi_{\rm H}}, 1, \ldots, 1)$. Only two matrix elements in the system unitary $U(x)$ will be necessary to calculate the CFI in this case; these elements can be evaluated as
\begin{eqnarray}
	\label{eq:ReceiverUnitary1} \hspace*{-.5cm}U_{1,1}(x)&=&\frac{1}{2}e^{i\phi_{\rm H}}\bigg[\frac{1}{M}\sum_{m=1}^Me^{i[\theta_m(x)-\theta_m(x_0)]}+1\bigg]\\
	\label{eq:ReceiverUnitaryM+1} \hspace*{-.5cm}U_{1,M+1}(x)&=&\frac{1}{2}e^{i\phi_{\rm H}}\bigg[\frac{1}{M}\sum_{m=1}^Me^{i[\theta_m(x)-\theta_m(x_0)]}-1\bigg].
\end{eqnarray}

Since $|\psi_0\rangle$ and $U(x)$ are Gaussian, the output of a real-quadrature homodyne measurement on mode $\hat{b}_1$ is characterized by the mean $d_{\rm H,1}(x)$ and variance $V_{\rm H,1,1}(x)$ of the first mode of the output state $\rho_{\rm H}$~\cite{Weedbrook2012,Ferraro2005}. These can be read off from the first moment vector $\vec{d}_{\rm H}(x)=S(x)\vec{d}_{0}$ and CM $V_{\rm H}(x)=S(x)V_0 S^T(x)$ of the state $\ket{\psi_{\rm H}}$, where the displacement vector and CM of the input state are given by Eqs.~\eqref{eq:d0} and \eqref{eq:V0} and the symplectic matrix $S(x)$ can be found from $U(x)$ using Eq.~\eqref{eq:symplectic}. Recognizing the optimality of $\alpha\in\mathbb{R}$ for the QFI, we set $\alpha=q_0$ (i.e., $p_0=0$). Therefore, the real quadrature of mode $\hat{a}_{M+1}$ has the only non-zero first moment among the input quadratures, and we have 
\begin{equation}
	\begin{split}
		d_{\rm H,1}(x)=&S_{1,M+1}(x)d_{0,M+1}\\
		=&\frac{1}{\sqrt{2}}S_{1,M+1}(x)\alpha,
	\end{split} 
	\label{eq:displacement_homodyne_lossless}
\end{equation}
where $S_{1,M+1}(x)=\Re\{U_{1,M+1}(x)\}$. To evolve the CM of the probe state [Eq.~\eqref{eq:V0}], 
%we recognize that the quadrature variances of the uncorrelated input modes $\hat{a}_m$ are all equal except for those of mode $\hat{a}_1$, so the input CM has the form $V_0=\frac{1}{2}[\mathcal{I}_{4M\cross 4M}+(2V_{0,1,1}-1)e_{1,1}+(2V_{0,2M+1,2M+1}-1)e_{2M+1,2M+1}]$, where  $e_{m,l}$ is a $4M\cross 4M$ matrix with the $\{m,l\}$ element equal to one and all other elements equal to zero. 
we use the fact $S(x)S^T(x)=\mathcal{I}$ for the symplectic matrix of any passive unitary to find that the variance of the measured quadrature is
\begin{equation}
	\begin{split}
		V_{\rm H,1,1}(x)=&\frac{1}{2}\big[1+S_{1,1}(x)^2(2V_{0,1,1}-1)\\
		&+S_{1,2M+1}(x)^2(2V_{0,2M+1,2M+1}-1)\big]\\
		=&\frac{1}{2}\big[1+S_{1,1}(x)^2(e^{2r}-1)\\
		&+S_{1,2M+1}(x)^2(e^{-2r}-1)\big],
	\end{split}
	\label{eq:CM_homodyne_lossless}
\end{equation}
where from Eq.~\eqref{eq:symplectic} we have $S_{1,1}(x)=\Re\{U_{1,1}(x)\}$ and $S_{1,2M+1}(x)=\Im\{U_{1,1}(x)\}$.

We now consider symmetric pure loss acting on the optical modes of Fig. \ref{fig:SensorDiagram}. We show in Appendix~\ref{apx:LossCommutation} that a symmetric pure loss channel across all modes of a system can be commuted with any passive unitary transformation. As a result, we can conceptually commute the pure loss channels to the very end of the optical circuit just before the homodyne measurement, such that only the effect of loss on the mode $\hat{b}_1$ will affect the CFI. Taking into account a transmittance of $\tau\leq1$ and using Eqs.~\eqref{eq:LossDisplacement} and \eqref{eq:LossCovariance}, the displacement vector and covariance matrix of the measured mode become
\begin{equation}
	d_{\rm H,1}(x;\tau)=\frac{1}{\sqrt{2}}\tau S_{1,M+1}(x)\alpha,  
	\label{eq:displacement_homodyne_lossy}
\end{equation}
and
\begin{equation}
	\begin{split}
		V_{\rm H,1,1}(x;\tau)=&\frac{1}{2}\big[1+\tau S_{1,1}(x)^2(e^{2r}-1)\\
		&+\tau S_{1,2M+1}(x)^2(e^{-2r}-1)\big],
	\end{split}
	\label{eq:CM_homodyne_lossy}
\end{equation}

For estimating a parameter $x$ from a Gaussian random variable with mean $d_{\rm H,1}(x;\tau)$ and variance $V_{\rm H,1,1}(x;\tau)$, the CFI is known to be given by $I_x=I_{x,\rm d}+I_{x,\rm V}$, where the two terms are given by
$I_{x,\rm d}=\left[\partial_x d_{\rm H,1}(x;\tau)\right]^2/V_{\rm H,1,1}(x;\tau)$ and
$I_{x,\rm V}=(1/2)\left[\partial_x V_{\rm H,1,1}(x;\tau)/V_{\rm H,1,1}(x;\tau)\right]^2.$ The CFI evaluated at $x=x_0$ is
\begin{eqnarray}
	\hspace{-.7cm}\label{eq:Ixd} I_{x_0,\rm d} &=& \frac{\tau\expval*{\partial\theta(x_0)}^2\sin^2(\phi_{\rm H})\abs{\alpha}^2}{1-\tau+\tau[\cos^2(\phi_{\rm H})e^{2r}+\sin^2(\phi_{\rm H})e^{-2r}]}\\
	\hspace{-.7cm}\label{eq:IxV} I_{x_0,\rm V} &=& \frac{\tau^2\expval*{\partial\theta(x_0)}^2\sin^2(2\phi_{\rm H})\sinh(2r)^2}{2\big(1-\tau+\tau[\cos^2(\phi_{\rm H})e^{2r}+\sin^2(\phi_{\rm H})e^{-2r}]\big)^2}.
\end{eqnarray}

The phase $\phi_{\rm H}$ is a free parameter for the receiver design, and three modes of operation can be identified. First, setting $\phi_{\rm H}=\pi/2$ maximizes the sensitivity of $d_{\rm H,1}(x;\tau)$ to changes in the parameter $x$, such that 
\begin{eqnarray}
	\label{eq:CFId1} I_{x_0,d}^{(1)}&=&\frac{\tau\expval*{\partial\theta(x_0)}^2\abs{\alpha}^2}{\tau e^{-2r} +1-\tau}\\
	\label{eq:CFIV1} I_{x_0,V}^{(1)}&=&0.
\end{eqnarray} 
This result is consistent with previously reported works on distributed sensing that involve a coherent state and SV probe and uniform loss but only consider equal phases on the $M$ modes \cite{Zhuang2018-zu,Grace2020}. This modality is preferred in the realistic scenario where the photon-unit energy of the laser source dominates that of the squeezed light and also has the advantage of a fixed homodyning angle that does not depend on the characterization of the system. In particular, in the absence of loss ($\tau=1$) we have
\begin{equation}
	I_{x_0}^{(1)}=\expval*{\partial\theta(x_0)}^2N_v(\sqrt{N_s}+\sqrt{N_s+1})^2,
	\label{eq:CFIresult}
\end{equation}
which exhibits Heisenberg scaling with probe energy. Second, the sensitivity of the covariance matrix $V_{\rm H,1,1}(x;\tau)$ can be optimized, yielding the optimal phase $\phi_{\rm H}=(1/2)\arccos(\tau \sinh(2r)/[1+2\tau\sinh(r)^2])$. In this case, the CFI becomes 
\begin{eqnarray}
	\label{eq:CFId2} I_{x_0,d}^{(2)}&=&\frac{1}{2}\frac{\tau\expval*{\partial\theta(x_0)}^2\abs{\alpha}^2}{\tau e^{-2r} +1-\tau}\\
	\label{eq:CFIV2} I_{x_0,V}^{(2)}&=&\frac{1}{2}\frac{\tau^2\expval*{\partial\theta(x_0)}^2\sinh^2(2r)}{1+4 \tau(1-\tau) \sinh^2(r)}.
\end{eqnarray}
Eq.~\eqref{eq:CFIV2} represents the best Fisher information that can be achieved with solely a squeezed vacuum input; in the absence of loss, this Heisenberg-limited term takes the form
\begin{equation}
	I_{x_0,V}^{(2)}=2\expval*{\partial\theta(x_0)}^2N_s(N_s+1),
	\label{eq:CFIV2lossless}
\end{equation}
which resembles the optimal performance achievable with Gaussian probes \cite{Matsubara2019}. Finally, the total CFI can be optimized by maximizing Eqs. \eqref{eq:Ixd} and \eqref{eq:IxV} with respect to $s\equiv\sin^2(\phi_H)$, which yields the optimal value
\begin{equation}
	s_{\rm opt}=\frac{[1+\tau(e^{2r}-1)][\tau\sinh(2r)+g]}{2\tau\sinh(2r)[1-\tau+\tau\cosh(2r)+g]},
	\label{eq:sopt}
\end{equation} 
where $g=\frac{1}{2}\abs{\alpha}^2[\csch(2r)+\tau\tanh(r)+\tau]$. This optimization gives
\begin{eqnarray}
	\nonumber \tilde{I}_{x_0,d}^{(3)}&=&\frac{\expval*{\partial\theta(x_0)}^2}{4\sinh^2(2r)[1-(1-e^{-2r})\tau]}\\
	\nonumber &&\times\{ 2\abs{\alpha}^2\tau \sinh^2(2r)\\
	\label{eq:CFId3}&&+\abs{\alpha}^4[1-(1-e^{2r})\tau]\}\\
	\nonumber 	 \tilde{I}_{x_0,V}^{(3)}&=&\frac{\expval*{\partial\theta(x_0)}^2}{8[1+4\tau(1-\tau)\sinh^2(r)]}\\
	\nonumber	&&\times \{4\tau^2\sinh^2(2r)\\
	\label{eq:CFIV3}	&&-\abs{\alpha}^4[1-(1-e^{2r})\tau]^2\csch^2(2r)\},
\end{eqnarray}
which reduces to
\begin{equation}
	\tilde{I}_{x_0}^{(3)}=\frac{\big\{2\tau\sinh^2(2r)+\abs{\alpha}^2[1-(1-e^{-2r})\tau]\big\}^2\expval*{\partial\theta(x_0)}^2}{8\sinh^2(2r)\big[1+4\tau(1-\tau)\sinh^2(r)\big]}.
	\label{eq:CFI3}
\end{equation}
Clearly, $0\leq s \leq 1$ for any operationally valid homodyne phase $\phi_{\rm H}$, and since it can be shown that $I_{x_0}$ is monotonically increasing with respect to $0\leq s \leq s_{\rm opt}$, the optimal receiver configuration uses $s=s_{\rm opt}$ when $s_{\rm opt}<1$ and $s=1$ (i.e., $\phi_{\rm H}=\pi/2$) when $s_{\rm opt}\geq 1$. The resulting CFI is 
\begin{equation}
	I_{x_0}^{(3)}=
	\begin{cases}
		\tilde{I}_{x_0}^{(3)},& \, s_{\rm opt}<1
		\\
		I_{x_0}^{(1)},& \, \textrm{otherwise}.
	\end{cases}
	\label{eq:CFI3cases}
\end{equation}
In the lossless case, we have
\begin{equation}
	\tilde{I}_{x_0}^{(3)}=\frac{\expval*{\partial\theta(x_0)}^2[8N_s(N_s+1)+N_v(\sqrt{N_s}-\sqrt{N_s+1})^2]^2}{32N_s(N_s+1)},
	\label{eq:CFI3Lossless}
\end{equation}
and $I_{x_0}^{(1)}$ is given by Eq.~\eqref{eq:CFIresult}.
%Equations \eqref{eq:CFId3} and \eqref{eq:CFIV3} are obtained by maximizing Eqs. \eqref{eq:Ixd} and \eqref{eq:IxV} with respect to $s\equiv\sin^2\phi_H$. Therefore we have the extra constraint $s\leq 1$ that can be solved for $r$ yielding the constraint $r\geq R(\tau,\alpha,\alpha^*)$, where $R(\tau,\alpha,\alpha^*)$ can be evaluated numerically. Any optimization over $r$ of the optimized over $\phi_H$ CFI, must take into account the extra condition $r\geq R(\tau,\alpha,\alpha^*)$. Also, the sanity check of getting CFI equal to zero for $\tau=0$, must take into account the valid values of $r$, i.e., $r\geq \lim_{\tau\rightarrow 0}R(\tau,\alpha,\alpha^*)$, which for this case results to $\sinh^2 r=(\textrm{total input energy})$, in which case said sanity check is verified.

%For the functional forms of $\vec{\theta}(x)$ that appear in applications we discuss below, we also show that the entangled probe across $M$ sensors yields $I_x = O(M^2)$, a Heisenberg scaling over $M$, which is not possible to attain with $M$ individual quantum sensors probing each phase, followed by classical postprocessing. This is important in applications where $M$ is large, e.g., a large sensor array in an RF photonic receiver antenna~\cite{Xia2019}, or a large number of orthogonal spatio-temporal modes of a near-field AFM probe.

\begin{figure}[tbp]
	\centering
	\includegraphics[width=\linewidth]{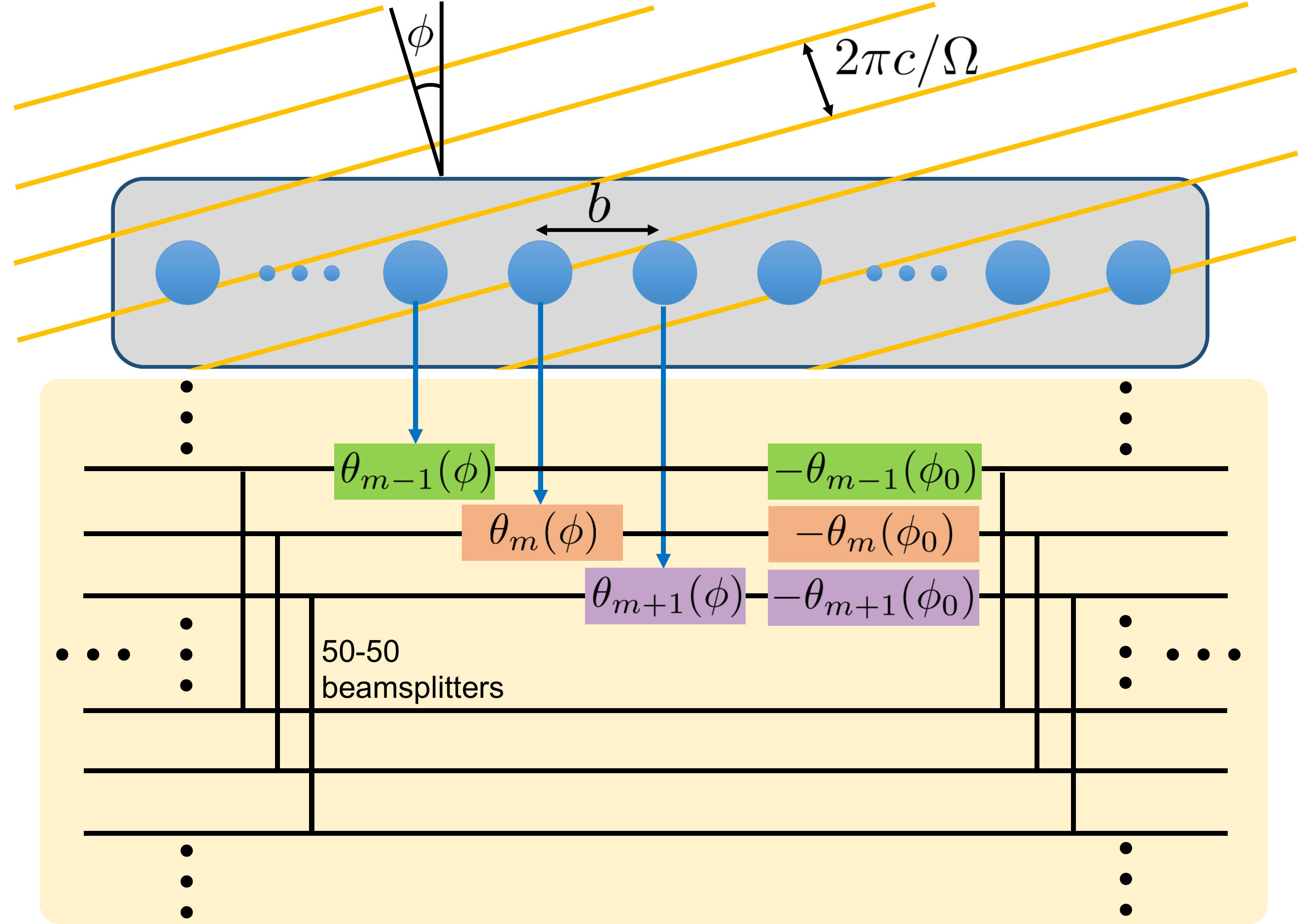}
	\caption{Application of our entanglement-enhanced sensor framework to a 1D phased array of RF-photonic sensors. The angle of incidence $\phi$ of the RF field (yellow lines) is estimated using $M$ RF-amplitude dependent optical phase modulators at lateral positions $mb$, which each modulate one arm of an optical MZI. Additional unitary transformations before and after the MZIs are not shown here (Fig.~\ref{fig:SensorDiagram}).}
	\label{fig:RF_diagram}
\end{figure}

\section{Optical Sensing Applications}
Our framework applies to any situation in which the phases of $M$ orthogonal optical modes are modulated by a common physical signal, one unknown parameter of which is to be estimated.
\begin{figure*}[tbp]
	\centering
	\includegraphics[width=\linewidth]{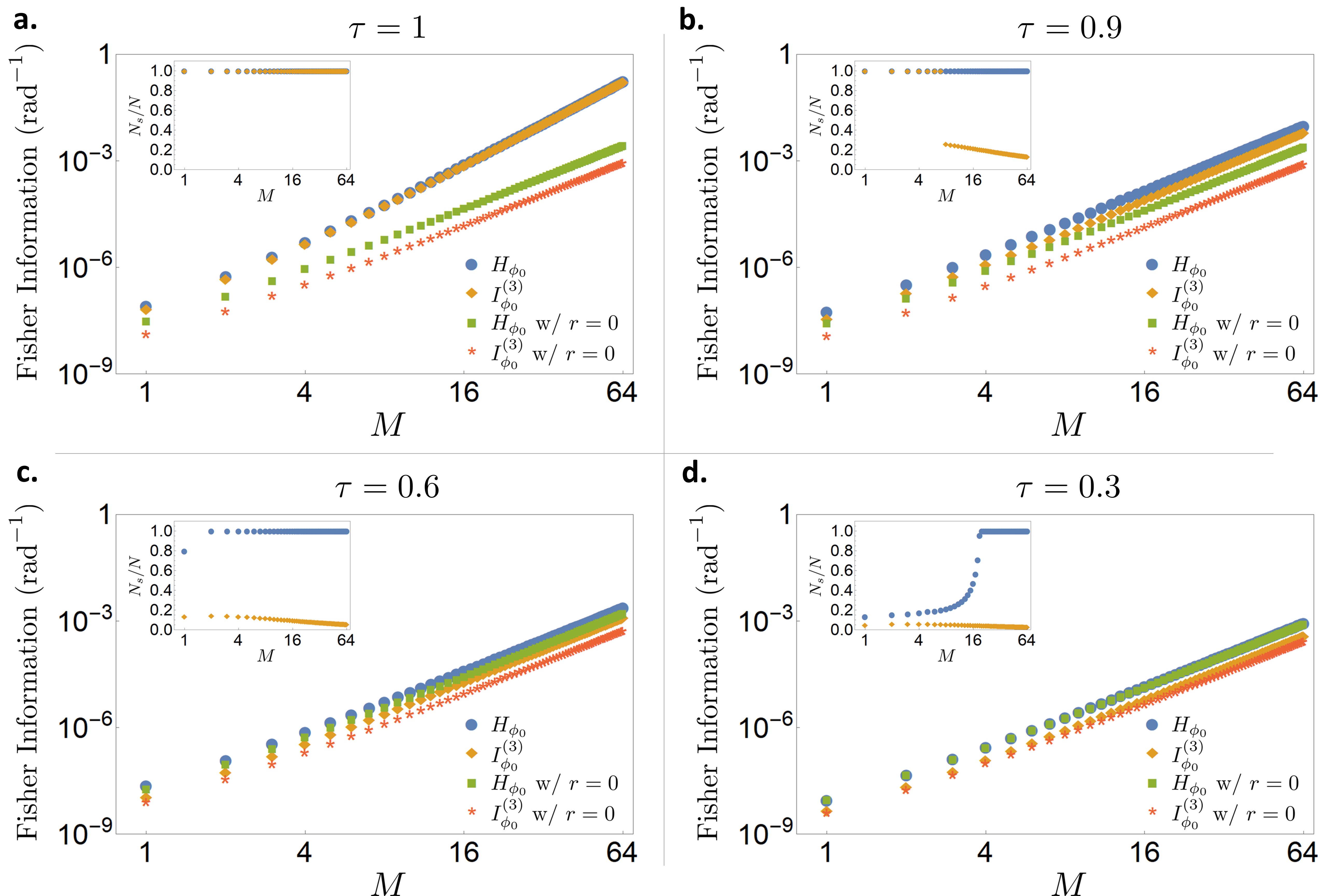}
	\caption{QFI and CFI for our entanglement-enhanced RF-photonic sensor as a function of the number of phases $M$ that encode the unknown parameter.  For fair comparison, we set the input energy to be proportional to $M$, i.e., $N=M \sinh^2 1$, resulting in a maximum of $7.69$ dB of single-mode squeezing when $M=64$ and $\alpha=0$. For the results shown for $H_{\phi_0}$ (blue circles) and $I_{\phi_0}^{(3)}$ (yellow diamonds), we optimize over input energy allocations for squeezing and displacement such that each point corresponds to a different numerically determined squeezing parameter. In the remaining two cases all energy is designated to the coherent state, such that $r=0$. For all plots we set $A=0.1$, $\Omega=30$ kHz, $b=10$ m, and $\phi_0=0$ in Eq. \eqref{eq:RF}. Insets: optimal fraction of total probe energy allocated to squeezed vacuum to maximize $H_{\phi_0}$ (blue circles) and $I_{\phi_0}^{(3)}$ (yellow diamonds).}
	\label{fig:RF_FI}
\end{figure*}
\subsection{RF Signal Estimation with a Photonic Receiver}
One example is the estimation of the angle of incidence $x\equiv\phi$ of an incident RF field using an $M$-pixel sensor array in an RF-photonic receiver antenna (Fig.~\ref{fig:RF_diagram}), an application for which CV entanglement was recently shown to improve upon classical estimation precision~\cite{Xia2020}. 

%$M$ RF-amplitude dependent optical phase modulators drive phases $\theta_m(\phi)$ on one branch of $M$ optical MZIs. In this case, the variation in optical phases arises from the spatial distribution of the sensors, which we take to have uniform linear separation $b$. For a freely time-evolving optical field $E_m(t)=Ee^{i\omega t+\theta_m(\phi)}$ with amplitude $E$ and optical frequency $\omega$, the position-dependent RF-modulated phase is \cite{Xia2019}
Each phase element in Fig.~\ref{fig:RF_diagram} is optically read out by an integrated-photonic MZI circuit. The optical-frequency continuous-wave (cw) field in the waveguide mode in one arm of the $m^{\textrm{th}}$ MZI is $E_m(t)=Ee^{i[\omega t+\theta_m(\phi)]}$, with
\begin{equation}
	\theta_m(\phi)=A\sin\bigg(\Omega\Big(t+\frac{mb\sin(\phi)}{c}\Big)\bigg),
	\label{eq:RF}
\end{equation} 
where $A$ is the RF-photonic amplitude-phase modulation efficiency, $\Omega$ is the center-frequency of the RF field, $mb$ is the relative position of the $m^{\textrm{th}}$ sensor, and $c=3 \times 10^8$ m/s is the speed of light. We assume $\abs{\phi-\phi_0}\ll1$, e.g., when the RF wave is known to arrive at close to normal incidence ($\phi_0=0$). Eq.~\eqref{eq:RF} can be used to compute the prefactors $\expval*{\partial\theta(\phi)}$ and $\expval*{\partial\theta(\phi)^2}$ in the QFI and CFI calculations.

Fig.~\ref{fig:RF_FI} plots the CFI $I_{\phi_0}^{(3)}$ for our fully optimized sensor design compared with the QFI $H_{\phi_0}$ evaluated at $\phi=\phi_0$, given total probe energy $N \propto M$ and varying degrees of symmetric loss. In each case, the energy allocation between the SV and coherent probe states is optimized to maximize the QFI or CFI (see insets). The QFI and CFI for classical sensors, where $r=0$, are also shown for reference. In the absence of loss (Fig.~\ref{fig:RF_FI}a), the Heisenberg scaling $H_{\phi_0}\propto I_{\phi_0}^{(3)} =O(N^2)$ observed in Eqs.~\eqref{eq:QFI_result} and \eqref{eq:CFI3Lossless} translates into a Heisenberg scaling $H_{\phi_0}\propto I_{\phi_0}^{(3)} = O(M^2)$ \cite{Zhuang2018-zu}. The $O(M^2)$ scaling of the CFI indicates that the receiver makes use of extra resources, i.e., additional distributed sensors with constant probe energy per sensor, to drive down the estimation error more efficiently than a classical receiver. 

This scaling advantage is only accessed by the use of the quantum SV probe; indeed, in the absence of loss, it is optimal to allocate all of the input energy to the SV state, in which case $I_{\phi_0}^{(3)}=I_{\phi_0}^{(2)}$. With $N=N_s$, in the high-energy limit ($N\gg1$) we find
\begin{equation} 						
	\lim_{N\to\infty}H_{\phi_0}=\lim_{N\to\infty}I_{\phi_0}^{(3)}=2\expval*{\partial\theta(\phi_0)}^2N_s(N_s+1),
	\label{eq:Limit}
\end{equation}
indicating that our receiver asymptotically achieves the quantum limit for distributed angle-of-incidence estimation in the lossless case. Importantly, our entanglement-enhanced scheme enables an increasingly large advantage over classical designs for large RF sensor arrays.

\begin{figure}[tb]
	\centering
	\includegraphics[width=\linewidth]{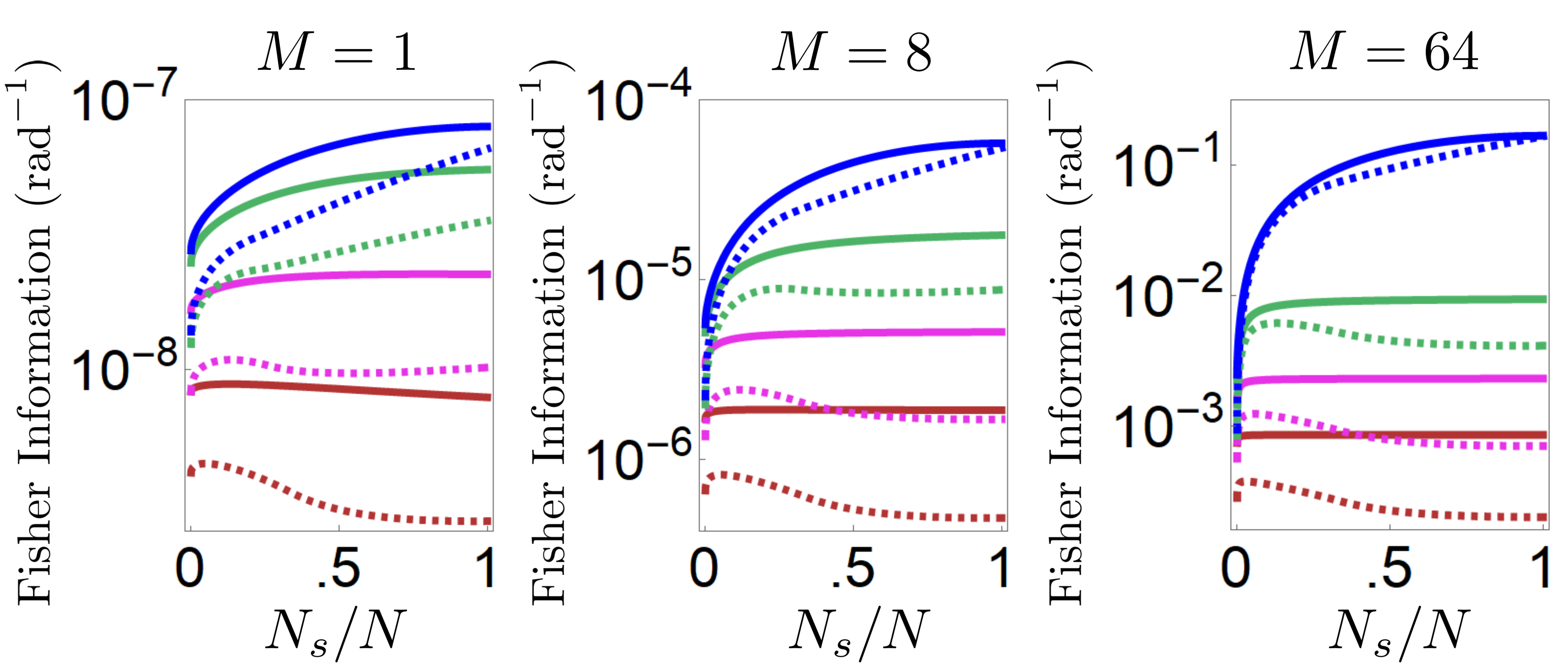}
	\caption{Fisher information as a function of energy allocation between coherent state and SV for the RF-photonic sensor application. Solid lines: QFI. Dashed lines: CFI. Colors (from high to low): Blue: $\tau=1$; Green: $\tau=0.9$; Magenta: $\tau=0.6$; Red: $\tau=0.3$.}
	\label{fig:RF_FI2}
\end{figure}

While the use of the complementary coherent state probe adds no benefit for our distributed sensor design in the absence of loss, it contributes to improved distributed sensing performance in the presence of loss. As loss is introduced (Figs.~\ref{fig:RF_FI}b-d), we observe several changes to the performance of our receiver. First, the Heisenberg scaling effect is immediately lost with the introduction of loss, which is a well-known phenomenon for quantum sensing.  On the other hand, the QFI of the classical sensor ($r=0$) draws nearer to the QFI of the entangled sensor as the severity of the loss increases, reflecting the fact that energy is increasingly shifted away from the SV state to the coherent state in the optimization of the QFI $H_{\phi_0}$. The coherent state is even more quickly prioritized as loss increases in the optimization of the CFI $I_{\phi_0}^{(3)}$, which falls gradually further from the QFI $H_{\phi_0}$ as loss is increased, though it never becomes sub-optimal by more than a factor of $\sim5$. In the high loss regime, it is easy to see that $s_{opt}>1$, so $I_{\phi_0}^{(3)}=I_{\phi_0}^{(1)}$, and with most of the energy allocated to the coherent state the performance of the entangled sensor converges to that of the classical sensor. These results demonstrate the value of including a coherent state in the probe design for lossy entangled sensing, as does Fig.~\ref{fig:RF_FI2}, which compares the fully optimized CFI against the QFI as the energy allocation ratio is traded off between the SV and coherent state. Allocating energy to the coherent state allows our design to achieve optimal estimation precision, especially with higher loss and larger distributed sensing networks, and also allows for a practically feasible increase of probe energy. In the conclusion section we briefly describe how to design a structured receiver that exactly achieves the QFI in all parameter regimes. 

%For an entangled probe in the high energy limit, the maximum QFI (Eq.~\eqref{eq:QFI_result}) will exhibit $H_{\phi}\propto2N^2$~\cite{Matsubara2019}. Our sensor framework can fully saturate this optimal QFI by eliminating the coherent state (i.e., $N_s=N$, $N_v=0$) and setting $\phi_H=(1/2)\cos^{-1}\big(-2\sqrt{N_s(1+N_s)}/(1+2N_s)\big)$, which achieves $I_{\phi}\propto 2N^2$. However, our practical configuration using a probe that employs both coherent and SV states, with $N_s=N^2/(2N+1)$, $N_v = N - N_s$ and $\phi_H=\pi/2$, retains a Heisenberg limited CFI $I_{\phi}\propto N^2$ that comes within a factor of two of the optimal QFI. This is a small price to pay in estimation precision in order to both reduce the required squeezed-light energy by at least two-fold and to use a fixed $\phi_H$ phase, which removes the concern that intensity fluctuations in the pump laser will negatively influence homodyne precision~\cite{Gramegna2020b}.

%\begin{figure}[tbp]
%	\centering
%	\includegraphics[width=0.9\columnwidth]{RF_results.pdf}
%	\caption{QFI and CFI for our entanglement-enhanced RF-photonic sensor, a sensor with product-state SV injected into each MZI, and a classical sensor with only coherent states ($A=0.1$, $\Omega=30$ kHz, $b=10$ m, $N=10M$, and $\phi_0=0$).}
%	\label{fig:RF_results}
%\end{figure}

\subsection{Optical Beam Displacement Tracking}
Our framework allows us to reproduce results from sensing tasks that can be unraveled into $M$ MZIs. One example is the use of spatial-mode entanglement to estimate a small lateral displacement $\delta$ of an optical beam \cite{Qi2018}, for example to implement a quantum-enhanced AFM. The input modes $\hat{a}_m$ are a set of $2M$ spatially overlapping, mutually-orthogonal (e.g., Hermite-Gauss) modes with negligible loss in the near-field regime. The output modes $\hat{b}_m$ are vacuum-propagation normal modes extracted by the receiver. Beam displacement causes modal-energy crosstalk quantified by a matrix $\Gamma$. This $2M$-mode crosstalk can be unitarily converted into a set of $M$ MZIs with phases $2\lambda_m \delta$, where the $\lambda_m$ depend on the eigenvalues of $\Gamma$. We thus can find the CFI evaluated at $\delta_0=0$, using Eq.~\eqref{eq:CFIresult}, where the prefactor becomes
\begin{equation}
	\expval*{\partial\theta(\delta_0)}^2=\frac{1}{M^2}\bigg(\sum_{m=1}^M 2\lambda_m\bigg)^2.
	\label{eq:beamDisplacement}
\end{equation}
Notably, since it was proven that $\sum_{m=1}^M \lambda_m\propto M^{3/2}$~\cite{Qi2018}, our analysis recovers the linear dependence of the CFI prefactor on $M$, which arises because the spatial-mode crosstalk becomes progressively more sensitive to the beam displacement $\delta$ as the mode order $m$ increases~\cite{Qi2018}. 

\subsection{Optical Temperature Gradiometry}
Our framework is also equipped to describe temporally entangled optical probes and dynamic physical systems, for which the time-bandwidth product $M=WT$ gives the number of orthogonal temporal modes for an optical source bandwidth $W$ and integration time $T$. For example, consider the estimation of the thermal conductivity $k$ of a uniform, dielectric rod with density $\rho$ and specific heat $c_p$, where the assumption $\abs{k-k_0}\ll1$ could stem from knowledge that $k$ diverges slightly from that of a known material due to physical impurities. We embed one branch of an optical fiber-based MZI at a position $y=y_0$ along the rod. If the rod is heated to an initial temperature distribution $u(y,0)$ and allowed to relax to steady-state, the sensor could probe the $k$-dependent optical phases induced by the temperature $u(y_0,t)$ at times $t=m/W$ using an $M$-temporal-mode-entangled CV state and $M$ (product) coherent states. The functional form $\theta_m(k)$ of the phases in the $M$ effective MZIs will depend both on the solution to the heat equation $\partial_t u(y,t) = (k/\rho  c_p) \partial^2_y u(y,t)$ and on the temperature-dependent Sellmeier equation for the optical refractive index of the fiber material. As long as the first derivatives $\partial_k\theta_m(k)$ can be computed analytically or numerically, Eq.~\eqref{eq:CFI3cases} gives the fully optimized CFI for the temporally entangled sensor, which can be compared with the QFI of Eq.~\eqref{eq:QFI_Loss} in the presence of fiber loss. In addition, with constant-power sources, the photon-unit probe energy $N$ will naturally scale linearly with $M$. The Heisenberg scaling $I_k=O(N^2)$ therefore extends to $I_k=O(M^2)$ under a lossless approximation. For low to moderate levels of loss, we expect a significant advantage from entanglement for long integration times (Fig.~\ref{fig:RF_FI}b-d).

\section{Discussion}
Many photonic sensing tasks can be reduced to estimating a scalar parameter $x$ that modulates phases ${\theta}_m(x)$ in $M$ MZIs. We discussed examples which include passive sensors whose receivers use quantum-enhanced computing, e.g., a quantum-enhanced RF photonic receiver, as well as active sensors that probe a scene with a non-classical illumination, e.g., beam tracking for AFM. We proved a Heisenberg limited scaling of the Fisher Information $I_x = O(N^2)$ in estimating the parameter in terms of the total photon-unit energy $N$ employed by the sensor. Furthermore, we argued that under certain circumstances, we see Heisenberg scaling in $M$ as well, i.e., $I_x = O(M^2)$. The latter, which is true for two of our examples, becomes significant when $M$ is large.

Additional constant factor improvements in $I_x$ are possible over our simple receiver design, e.g., by optimizing $U_I$ for known functions $\theta_m(x)$. This will involve arbitrary tuning of $\mathcal{O}(M^2)$ phases in a programmable linear-optic circuit~\cite{Reck1994-to,Clements2016-jr}. Under optical loss and noise, the Heisenberg  scalings with respect to $N$ and $M$ will disappear, and $I_x = O(MN)$ will prevail. However, there will be a constant factor improvement in $I_x$ over a classical sensor in the long integration time limit, which can be significant for moderate losses and if $M$ is large~\cite{Zhuang2018-zu,Oh2020,Grace2020}. %We leave the above to future work, as well as analysis of other applications in quantum process verification for photonic quantum circuits, quantum network tomography, and quantum estimation problems.

\begin{figure}[tb]
	\centering
	\includegraphics[width=\linewidth]{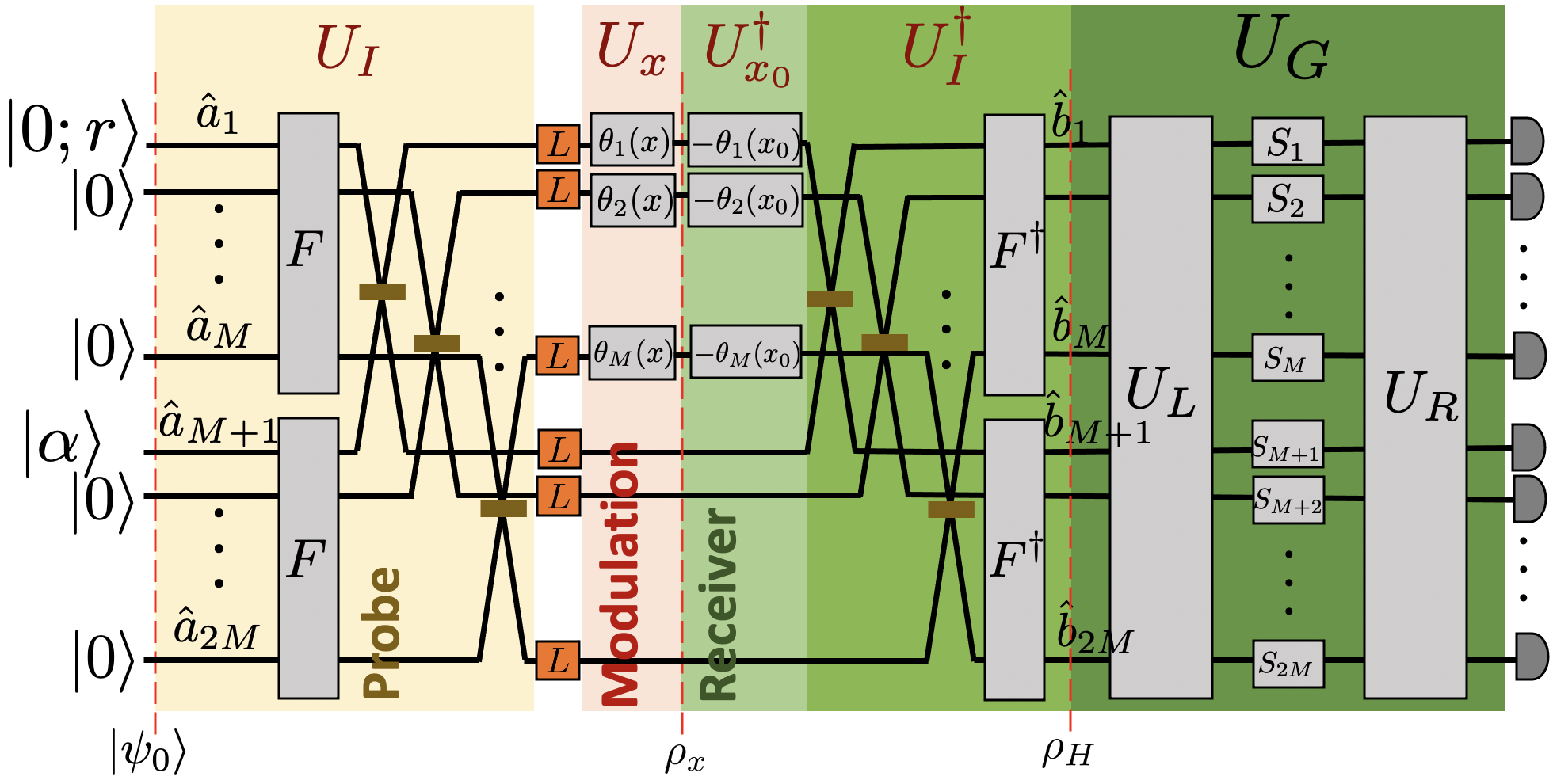}
	\caption{A schematic of the optimal structured receiver design that attains the QFI for estimating the single scalar parameter $x$ for the problem studied in this paper, as sketched in Fig.~\ref{fig:SensorDiagram}. $U_L$ and $U_R$ are multimode passive linear-optical transformations, and $S_1, \ldots, S_{2M}$ are single-mode squeezers. The homodyne detection receiver strategy depicted in Fig.~\ref{fig:SensorDiagram} is strictly suboptimal in the presence of losses.}
	\label{fig:lossyreceiverdesign}
\end{figure}
We note that if one desires to find the optimal receiver, a tractable method could be based on the symmetric logarithmic derivative (SLD) whose eigenvectors provide an optimal measurement. Indeed, the SLD is a Gaussian polynomial \cite{Jiang2014,Bakmou2020} (i.e. at most quadratic in $\hat{a}_i$ and $\hat{a}^\dagger_i$ or equivalently in $\hat{q}_i$ and $\hat{p}_i$, where index $i$ counts the modes). Therefore, one could follow standard methods to diagonalize the Gibbs matrix corresponding to the SLD operators, i.e., similarly to \cite{Oh2019} where single mode systems have been examined. If the SLD operators for each parameter commute \cite{Ragy2016} so that there exists a measurement which attains the QFI, a similar method to \cite{Oh2020} could work in a multi-parameter and multi-mode setting potentially revealing a rich measurement structure. For estimating a single scalar parameter embedded in a multimode Gaussian state, which is relevant to the problem studied in this paper, the optimal receiver design takes the form shown in Fig.~\ref{fig:lossyreceiverdesign}. The optimal receiver takes the form of a multi-mode Gaussian unitary transformation of the modulated lossy multimode Gaussian state $\rho_H$ followed by mode-resolved photon number resolving (PNR) on {\em all} $2M$ modes. The general Gaussian unitary $U_G$ preceding the PNR detector array can be further decomposed into a linear optical unitary $U_L$ followed by single-mode squeezers $S_1, \ldots, S_{2M}$, and another linear optical unitary $U_R$, as shown~\cite{Braunstein2005-se}. We will explore the optimal receiver design problem for general Gaussian multi-parameter estimation problems in future work.

{\em Acknowledgement}---This research was supported by ONR contract no. N00014-19-1-2189, and USRA under a NASA contract. We acknowledge valuable discussions with Quntao Zhuang, Zheshen Zhang, and Linran Fan. MRG and CNG contributed equally to this work.

\appendix
\section{QFI Calculation with Pure Uniform Loss}
\label{apx:QFI-Loss}
To calculate the QFI for a parameter $x$ which is embedded in a set of phases $\{\theta_1(x),\ldots,\theta_M(x)\}$, we need first to calculate the QFI matrix for the multiple phase estimation problem whose matrix elements $H_{kl}$ are,
\begin{equation}
	\begin{split}
		H_{kl}=&\text{Re}\Bigg[\sum_{\vec{n}=0}^\infty \frac{(\partial_{\theta_k} E_{\vec{n} }) (\partial_{\theta_l} E_{\vec{n}}) } {E_{\vec{n}}}\\
		&+4 \sum_{\vec{n},\vec{m}=0}^\infty E_{\vec{m}} \left(\frac{E_{\vec{n}}-E_{\vec{m}}}{E_{\vec{n}}+E_{\vec{m}}}\right)^2  \langle e_{\vec{n}} | \hat{n}_k | e_{\vec{m}} \rangle \langle e_{\vec{m}} | \hat{n}_l | e_{\vec{n}} \rangle\Bigg]
	\end{split}
	\label{eq:QFI1}
\end{equation}
where $E_{\vec{n}}$ and $|e_{\vec{n}}\rangle$ are respectively the eigenvalues and eigenvectors of the $2M$-dimensional state $\rho_x$ just after modulation by the parameter-dependent phases. Then, we apply the Jacobian transformation and the QFI for parameter $x$ is given by Eq.~\eqref{eq:QFI_Jacobian}. Here we focus on calculating $H_{kl}$.

The final state (and the state in any stage of our setup) is Gaussian. Therefore, we can exploit the phase-space toolbox to find the eigenvalues and eigenvectors of the final state. We also note that since we assume uniform loss, we can commute the single-mode single channels with unitary transformations (see Appendix~\ref{apx:LossCommutation}). We choose for our derivation to conceptually commute the pure loss channels to act on the input state, i.e., before the interferometer. Also, the eigenvalues of the state will not change after the action of pure loss as subsequently the state is transformed under unitary operations.

The $4M\times4M$ CM $V_{\tau}$ of the state just after the pure loss channels is found by transforming the input CM $V_0$ [Eq.~\eqref{eq:V0}] according to Eq.~\eqref{eq:LossCovariance}. By calculating the (usual) eigenvalues of $|i \Omega V_{\tau}|$, we find that the symplectic diagonalization of $V_{\tau}$ is 
\begin{equation}
	V_{\tau}=S_{\tau} (D_{\tau}\oplus D_{\tau}) S_{\tau}^T,
	\label{eq:SympDiag}
\end{equation}
where $D_{\tau}=\text{diag}\left(\nu_1,1/2,\ldots,1/2\right)$, and where $\nu_1=\sqrt{\tau (1-\tau)\sinh^2 r+1/4}$ (we will examine the $S_{\tau}$ after we find the eigenvalues). We write the eigenvalues compactly as, 
\begin{eqnarray}
	\label{eq:eigenvalues}	E_{\vec{n}} = \frac{\bar{N}_1^{n_1}}{(\bar{N}_1+1)^{n_1+1}} \delta_{n_2,0} \ldots \delta_{n_{2M},0},
\end{eqnarray}
where $\bar{N}_1$ \eqref{eq:ThermalPhotons} is the mean thermal photon number in the first mode arising from the action of the pure loss channel on squeezed vacuum. As discussed before, Eq. \eqref{eq:eigenvalues} are the eigenvalues of $\rho_x$ as well. 

Now, let us find the eigenvectors of the final state. It is easy to verify the following equations,
\begin{eqnarray}
	\label{eq:SympMatrix}	S_{\tau}&=&\diag\left(e^{s},1\ldots,1,e^{-s},1,\ldots,1\right)\\
	\label{eq:Cond1}S_{\tau} \Omega S_{\tau}^T&=&\Omega\\
	\label{eq:Cond2}S_{\tau}^T \Omega S_{\tau}&=&\Omega\\
	\label{eq:Cond3} \text{Det}S_{\tau}&=&1,
\end{eqnarray}
where $S_{\tau}$ is a $4M\times 4M$ matrix. Equations \eqref{eq:Cond1}, \eqref{eq:Cond2}, and \eqref{eq:Cond3} guarantee that $S_{\tau}$ is a symplectic matrix and, as per Eqs. \eqref{eq:SympDiag} and \eqref{eq:SympMatrix}, the symplectic matrix $S_{\tau}$ diagonalizes (in the symplectic sense) the CM $V_{\tau}$. From the structure of $S_{\tau}$ we deduct that it corresponds to a single mode squeezer on the first mode and identity on the rest of the modes.
In the absence of loss (i.e., $\tau=1$), Eq. \eqref{eq:squeezing} gives $s=r$.

We must not forget the displacement of the state prior to loss in the $M+1$ mode. After loss is applied [Eq.~\eqref{eq:LossDisplacement}], the displacement of the state is given by $\vec{d}_{\tau}=(0,\dots,0,\sqrt{\tau}q_0,0,\ldots,0,\sqrt{\tau}p_0,0,\ldots 0)$, describing a phase-space displacement by $\sqrt{\tau}\alpha$ on the $M+1$ mode. Therefore, the diagonal form of final state is
%\begin{eqnarray}
%	U_0= D(\sqrt{\tau}\alpha)^{(M+1)} U_{\text{sq}}^{(1)}(s),
%\end{eqnarray}
%where the superscripts denote the mode each unitary acts on. The diagonalization of $\rho_0$ reads,
\begin{eqnarray}
	\rho_x = \sum_{\vec{n}} E_{\vec{n}} |e_{\vec{n}}\rangle \langle e_{\vec{n}} |,
\end{eqnarray}
where the eigenvalues are given in Eq. \eqref{eq:eigenvalues}, the eigenvectors $|e_{\vec{n}}\rangle$ are
\begin{eqnarray}
	\label{eq:eigenvector}	|e_{\vec{n}}\rangle=U_{x} U_I D(\sqrt{\tau}\alpha)^{(M+1)} U_{\text{sq}}^{(1)}(s) |\vec{n}\rangle,
\end{eqnarray}
%Knowing the eigenvalues given in Eq. \eqref{eq:eigenvalues}, the only thing remaining is to calculate $\langle e_{\vec{n}} | \hat{n}_k | e_{\vec{m}} \rangle$.
%Change of variables in Eq. \eqref{eq:QFI1} or by observing Eq. \eqref{eq:QFI2} (even though in \eqref{eq:QFI2} I have dropped the parameter dependence of the eigenvalue, it can be generalized to include parameter depended eigenvalues), I find that,
%\begin{eqnarray}
%	H= \sum_{l,k=1}^{2M} \partial_x \theta_k (x) \partial_x \theta_l (x) H_{lk}=(\partial_x \vec{\theta})^T\cdot H_{\vec{\theta}}\cdot(\partial_x \vec{\theta}),
%\end{eqnarray}
%where $H_{lk}$ is the QFI matrix (denoted $H_{\vec{\theta}}$) element for estimating the vector of unknown phases $\vec{\theta}=\left(\theta_1,\theta_2,\ldots\right)$.
%\textbf{Is $\bm{H_{lk}}$ for the Gaussian probe we use and pure loss known in the literature? I think it should be. If yes, then the calculation of $\bm{H}$ is straightforward.}
where $U_x$ is the unitary which imprints the phases onto the state, $U_I$ is the passive unitary interferometer, $ D(\sqrt{\tau}\alpha)^{(M+1)}$ is a displacement operator that acts only on mode $M+1$, $ U_{\text{sq}}^{(1)}(s)$ is a single-mode squeezer acting on the first mode, and $|\vec{n}\rangle=|n_1,\ldots ,n_{2M}\rangle$ represents a product of $2M$ Fock states.

Coming back to the QFI of Eq. \eqref{eq:QFI1} we observe that our eigenvalues [\eqref{eq:eigenvalues}] do not depend on the parameter, as expected since we our parameters are imprinted by a unitary operation. Therefore, Eq. \eqref{eq:QFI1} simplifies to
\begin{equation}
	\label{eq:QFI3}	H_{kl}=4 \sum_{\vec{n},\vec{m}=0}^\infty E_{\vec{m}} \left(\frac{E_{\vec{n}}-E_{\vec{m}}}{E_{\vec{n}}+E_{\vec{m}}}\right)^2  \langle e_{\vec{n}} | \hat{n}_k | e_{\vec{m}} \rangle \langle e_{\vec{m}} | \hat{n}_l | e_{\vec{n}} \rangle,
\end{equation}
where we took under consideration the fact that the QFI elements for our case are always real. Taking into account the form of Eq. \eqref{eq:eigenvalues}, Eq. \eqref{eq:QFI3} can be simplified further as the Kronecker deltas will set all Fock number to zero, except of $n_1$ and $n_{M+1}$. We get,
\begin{equation}
	\label{eq:QFI4}	
	\begin{split}
		H_{kl}=&4 \sum_{n_1,m_1=0}^\infty E_{m_1} \left[\left(\frac{E_{n_1}-E_{m_1}}{E_{n_1}+E_{m_1}}\right)^2-1\right]  A_{n_1,m_1}^{(k)} A_{m_1,n_1}^{(l)}\\
		&+4 \sum_{n_1=0}^{\infty} E_{n_1} B_{n_1}^{(k,l)},
	\end{split}
\end{equation}
where,
\begin{eqnarray}
	\label{eq:E1} \hspace*{-1cm}E_{n_1} &=& \frac{\bar{N}_1^{n_1}}{(\bar{N}_1+1)^{n_1+1}} \\
	\label{eq:A} \hspace*{-1cm} A_{n_1,m_1}^{(k)} &=& \langle \psi(n_1)|U^{(1)\dagger}_{\text{sq}}(s) U_I^\dagger n_k U_I U^{(1)}_{\text{sq}}(s) |\psi(m_1)\rangle\\
	\label{eq:B} \hspace*{-1cm}	B_{n_1}^{(k,l)} &=& \langle \psi(n_1)|U^{(1)\dagger}_{\text{sq}}(s) U_I^\dagger n_k n_l U_I U^{(1)}_{\text{sq}}(s) | \psi(n_1)\rangle
\end{eqnarray}
and where the state $|\psi(n)\rangle=| n,0,\ldots,0,\sqrt{\tau} \alpha,0,\ldots,0 \rangle$ is a product state between a Fock state $|n\rangle$ in mode 1, a coherent state $|\sqrt{\tau}\alpha\rangle$ in mode $M+1$, and vacua in the other $2M-2$ modes. For the specific $U_I$ considered in Eq.~\eqref{eq:UI}, working in the Heisenberg picture, and converging all summations involved in Eq. \eqref{eq:QFI4}, it is straightforward to calculate the amplitudes of Eqs. \eqref{eq:A} and \eqref{eq:B} to arrive at Eq.~\eqref{eq:QFI_phases}. Given a specific set of functions $\theta_m(x)$, one can straightforwardly apply Eq. \eqref{eq:QFI_Jacobian} to obtain the QFI for the parameter $x$.

\section{Commutation of a Symmetric Pure Loss Channel with Gaussian Operations}
\label{apx:LossCommutation}
Here we show that a symmetric set of $M$-mode pure loss channels with transmission coefficient $\tau$ on each mode commutes with any passive (i.e., energy-preserving) Gaussian unitary transformation when the state being acted upon is Gaussian. Consider an arbitrary $M$-mode Gaussian state with displacement vector $\vec{d}$ and covariance matrix $V$. An arbitrary passive unitary matrix $U_{\rm B}$, which can in general be decomposed into two-mode beamsplitters and phase shifts \cite{Reck1994}, is described by its symplectic matrix $S_{\rm B}$ from Eq.~\eqref{eq:symplectic} with $S_{\rm B}^TS_{\rm B}=\mathcal{I}$. In addition, if the state is subjected to symmetric pure loss, the resulting displacement vector $\vec{d}_{\tau}$ and CM $V_{\tau}$ are calculated according to Eq.~\eqref{eq:LossDisplacement} and Eq.~\eqref{eq:LossCovariance}. Clearly, $S_{\rm B}$ commutes with both $X_{\tau}$ and $Y_{\tau}$.

Acting the unitary $U_{\rm B}$ on the state at the output of the loss channel, we find a displacement vector
\begin{equation}
	S_{\rm B}\vec{d}_{\tau}=S_{\rm B}X_{\tau}\vec{d}=X_{\tau}S\vec{d}=X_{\tau}\vec{d}_{\rm B},
	\label{eq:pureLossProof1}
\end{equation}
where $\vec{d}_{\rm B}=S_{\rm B}\vec{d}$ is the displacement vector of the initial state transformed by $U_{\rm B}$. The corresponding covariance matrix is
\begin{equation}
	\begin{split}
		S_{\rm B}V_{\tau}S_{\rm B}^T=&S_{\rm B}X_{\tau}VX_{\tau}^TS_{\rm B}^T + S_{\rm B}Y_{\tau}S_{\rm B}^T\\
		=&X_{\tau}S_{\rm B}VS_{\rm B}^TX_{\tau}^T+S_{\rm B}S_{\rm B}^TY_{\tau}\\
		=&X_{\tau}V_{\rm B}X_{\tau}^T+Y_{\tau},
	\end{split}
	\label{eq:pureLossProof2}
\end{equation}
where $V_{\rm B} = S_{\rm B}VS_{\rm B}^T$ is the transformed covariance matrix of the initial state subjected to $U_{\rm B}$. Taken together, Eqs.~\eqref{eq:pureLossProof1} and \eqref{eq:pureLossProof2} prove the commutation of symmetric pure loss and passive Gaussian unitaries for Gaussian states.

\bibliographystyle{unsrt}
\bibliography{DistrFunctionBIB}
\end{document}